\documentclass[11pt]{article}
\usepackage[utf8]{inputenc}
\usepackage[]{geometry}

\usepackage{graphicx}
\usepackage{caption}
\usepackage{subcaption}
\usepackage[english]{babel}
\usepackage{csquotes}
\usepackage{amsmath}
\usepackage{amssymb}
\usepackage{bm}
\usepackage[section]{placeins}
\usepackage{nameref}
\usepackage{xcolor,colortbl}
\usepackage[hyphens]{url}
\usepackage[hidelinks]{hyperref}
\usepackage[numbers]{natbib}
\usepackage[ruled]{algorithm2e}
\usepackage{booktabs}
\usepackage{tabularray}
\usepackage{multirow}
\usepackage{mathpazo}

\title{On the Creativity of AI Agents}

\author{
Giorgio Franceschelli\\
University of Bologna\\
\texttt{giorgio.franceschelli@unibo.it}
\and
Mirco Musolesi\\
University College London and
University of Bologna\\
\texttt{m.musolesi@ucl.ac.uk}
}
\date{}
\begin{document}

\maketitle

\begin{abstract}
Large language models (LLMs), particularly when integrated into agentic systems, have demonstrated human- and even superhuman-level performance across multiple domains. Whether these systems can truly be considered creative, however, remains a matter of debate, as conclusions heavily depend on the definitions, evaluation methods, and specific use cases employed.
In this paper, we analyse creativity along two complementary macro-level perspectives. The first is a \textit{functionalist} perspective, focusing on the observable characteristics of creative outputs. The second is an \textit{ontological} perspective, emphasising the underlying processes, as well as the social and personal dimensions involved in creativity. We focus on LLM agents and we argue that they exhibit functionalist creativity, albeit not at its most sophisticated levels, while they continue to lack key aspects of ontological creativity. Finally, we discuss whether it is desirable for agentic systems to attain both forms of creativity, evaluating potential benefits and risks, and proposing pathways toward artificial creativity that can enhance human society.
\end{abstract}

\section{Introduction}


Recent years have seen the rise of agentic systems (e.g., \cite{geminiteam2025gemini,kimiteam2026kimi,openai2025o3}), i.e., autonomous or semi-autonomous software entities capable of reasoning, acting, and interacting with external environments\footnote{The study of (multi-agent) agent systems has a long tradition in Computer Science, which is often unfortunately overlooked (see, for example, \cite{shoham2008multiagent,wooldridge2009introduction}). Early agent systems were primarily based on logic, and as a result, their capabilities were inherently limited. However, these systems laid important foundations for the advances we see today.}. These systems can plan multi-step solutions, invoke tools, retrieve information, and adapt their behaviour based on feedback, enabling them to tackle complex and dynamic tasks. At the core of many such systems lie large language models (LLMs), whose recent advancements are revolutionising several fields \cite{bommasani2021opportunities} due to their remarkable emergent capabilities \cite{wei2022emergent}.
LLMs provide the underlying reasoning and language understanding that make agentic behaviour possible. In particular, they can solve complex and even previously unseen tasks through prompt engineering \cite{reynolds2021prompt} and in-context learning \cite{brown2020language}, i.e., by adjusting the input text or augmenting it with a few examples of expected outcomes. Building on these capabilities, modern agentic systems are typically structured around an often fine-tuned LLM\footnote{We envisage the rise of agents based on multimodal foundation models. While this article focuses on LLM agents, the ideas presented here naturally extend to this broader class of systems.} that serves as the basis for both \textit{acting} (by generating textual commands to invoke tools) and \textit{perceiving} (by incorporating retrieved information or environmental state into its input prompt).

These LLM agents have evolved beyond plain text generation into fully fledged acting systems capable of using tools and planning over complex tasks. For example, their capabilities have progressed from producing programming code in a single pass \cite{austin2021program} to orchestrating more sophisticated workflows, in which tasks are decomposed into smaller sub-problems and addressed iteratively. Within these workflows, systems can debug and optimise generated code based on unit test performance, evaluate it at runtime, and return a solution only once all specified requirements are satisfied \cite{hu2025quality}.

Their widespread use in tasks traditionally associated with creativity and innovation (not only in writing or coding, but also in academic research \cite{si2025can}, scientific discovery \cite{du2025accelerating}, and problem-solving \cite{tian2024macgyver}) has sparked intense debate over whether they can truly be considered creative. Depending on the definition of creativity assumed \cite{runco2025updating, wang2024ai}, the psychometric tests employed \cite{guzik2023originality, koivisto2023best}, the use cases considered \cite{marco2025small, ismayilzada2025evaluating}, and the level of proficiency of human evaluators \cite{porter2024aigenerated, davis2024chatpgt}, the research community is somewhat polarized between those arguing that LLMs and LLM agents possess (truly) creative capabilities and those who contest this view.

In this paper, we propose a dualistic framework that integrates both views, providing a coherent way to discuss creativity. The framework differentiates a \textit{functionalist approach}, focusing on the observable traits of artefacts and ideas, from an \textit{ontological approach}, which explores the essential nature of the creative process itself. Grounding our analysis in the mechanistic behaviour of LLM-based agentic systems, we explore how they can achieve specific forms of functionalist creativity, yet remain limited with respect to deeper, ontological aspects of creativity. We highlight the gaps in current agentic systems and consider whether autonomous agents might reach higher creative capacities, proposing research directions where enhancing creativity could simultaneously support societal benefits.

\section{From Language Models to AI Agentic Systems} \label{llm_agents}

Large Language Models (LLMs) are autoregressive generative models built on the Transformer architecture \cite{vaswani2017attention}. They are described as large because they contain billions of parameters, and as language models because they are trained on vast text corpora (e.g., almost the entirety of Web content) to learn the structure and patterns of human language.  Technically speaking, a $\boldsymbol{\theta}$-parametrised autoregressive language model is a probability distribution $p_{\boldsymbol{\theta}}(\mathbf{x})$ over a variable-length text sequence $\mathbf{x} = (x_1 \ldots x_T)$, where $T$ is the sequence length, and each token $x_t$ is in a finite vocabulary $\mathcal{V}$ of size $N$. A token can be a word, a sub-word, or even a single character, a punctuation mark, or any other symbol or combination of symbols that appears in written form. The probability distribution is factorized as $p_{\boldsymbol{\theta}}(\mathbf{x}) = \prod_{t=1}^T p_{\boldsymbol{\theta}}(x_t | \mathbf{x}_{<t})$, where $\mathbf{x}_{<t} = x_1 \ldots x_{t-1}$. The language model is usually trained to maximise the likelihood of the true distribution $p^*(\mathbf{x})$ for any $\mathbf{x}$ from a reference dataset (the training set). In other words, given an input $\mathbf{x}_{<t}$, the model learns to approximate the probability of each token from $\mathcal{V}$ being $x_{t}$. By sampling from these learned next-token probability distributions, we can use the LLM to generate new sentences. Given a conditional input (the prompt) $\mathbf{z} = (z_1 \ldots z_L)$, we can decode $p_{\boldsymbol{\theta}}(\mathbf{x}|\mathbf{z})$ as the continuation of $\mathbf{z}$, i.e., through the factorized representation $p_{\boldsymbol{\theta}}(\mathbf{x} | \mathbf{z}) = \prod_{t=1}^T p_{\boldsymbol{\theta}}(x_t | \mathbf{x_{<t}}, \mathbf{z})$. Commonly, after maximising the likelihood of the training set, the language model is subject to an additional, shorter training (called fine-tuning) with reinforcement learning. Here, the $\boldsymbol{\theta}$-parametrised model is incentivized (or discouraged) to generate completions $\mathbf{x}$ that maximize (or minimize) human feedback \cite{ouyang2022training} or other desired properties such as correctness, completeness, and appropriateness given the prompt $\mathbf{z}$ \cite{franceschelli2024reinforcement}.

In essence, given a fragment of text, what an LLM does is to predict what is likely to come next according to its model of the statistics of human language \cite{shanahan2024talking}, which is derived from training data and possibly tuned with human feedback. Then, the output returned to the user is sampled according to these predictions by means of one of several available sampling strategies, which might truncate certain tokens (e.g., \cite{fan2018hierarchical,holtzman2020curious,minh2025turning}) as well as sharpen \cite{peeperkorn2024temperature} or flatten \cite{song2025good} their likelihood but still leverage the predicted next-token probability distribution. However, LLMs are rarely used in isolation, particularly outside the research community. Instead, they are typically integrated into larger agentic systems, sometimes confusingly also referred to as ``large language models'', even though they are, strictly speaking, applications built around an LLM core \cite{shanahan2024still}.

LLM agents are \textit{systems} where LLMs receive input in natural language from their environment and take autonomous actions to accomplish specific tasks \cite{plaat2025agentic}. Their basic building block is an LLM enhanced with augmentations such as retrieval (i.e., the LLM can generate its own search query, and the results are included in the next-step prompt) \cite{lewis2020retrieval}, tools (i.e., the system can call tools by writing programming code, including predefined APIs, that is executed in a container and whose response is returned as part of the next-step prompt) \cite{qin2024toolllm}, and memory (i.e., the system stores and maintains an evolving memory state from prior interactions, and at each timestep the LLM input relies on additional, relevant information retrieved from such a memory in addition to the current observation and the original prompt) \cite{hu2026memory}.
LLM agents can differ in terms of memory, reasoning, and acting tools \cite{wang2024survey}. For instance, reasoning can happen due to a few-shot chain-of-thought \cite{wei2022chainofthought} or a zero-shot ``let's think step by step'' chain-of-thought \cite{kojima2022large}, and acting can even involve physical robots \cite{ahn2022icanisay}. Finally, agentic LLMs can interact not only with the user, but also with each other, based on specific roles (as in simulated companies \cite{qian2024chatdev,wu2025perhaps}) or even in open-ended settings where the LLM agents do not have predetermined roles and interact semi-spontaneously, forming in-silico societies \cite{park2023generative}.
Regardless of the tools and techniques they can dispose of, these systems can be built as \textit{workflows}, i.e., systems where LLMs and tools are orchestrated through predefined code paths and the task is decomposed a priori by the developer into fixed parallel or consecutive steps that involve LLMs in different ways; or as \textit{agents}, i.e., systems where LLMs dynamically direct their own processes and tool usage based on environmental feedback to accomplish given tasks \cite{schluntz2024building}.

In other words, LLM-based agentic systems are programs where LLM outputs control the workflow. While this might sound amazing (and it partially is), in practice, this is done by running the very same LLM inference we mentioned before in a loop: at each timestep, the input of the LLM is made of the user prompt, the last observation from the environment, and a summary of the results of previous iterations; and at each timestep, the output of the LLM is usually composed of two parts: first, the LLM ``reasoning" on what it should do (or, more correctly, predicting what is likely to come next after the prompt and some ``thinking'' keywords), and second, the LLM writing actions as executable code snippets that make use of one or more available tools (or, more correctly, predicting what is likely to come next after the prompt, the generated ``reasoning'', and some ``acting'' keywords). Then, the program merely executes it and appends its output to the LLM output as the ``observation'' from the environment. The final answer returned to the user is simply the last LLM output (which does not contain any code snippet, as it predicts it already has all necessary information) \cite{roucher2024introducing}. In this way, the LLM can solve tasks from a great variety of domains, provided that it has access to the appropriate tools, which definition plays a key role in the success of LLM-based agentic systems \cite{wu2025introducing}.

\section{The Present of Creative AI Agents: A Theory-Grounded Taxonomy}

The debate around creativity in AI is often fragmented and framed at different levels of abstraction. Opposite claims typically arise because there is no agreement upon what creativity means and how it should be evaluated in the context of AI and, particularly, agentic systems. Here, we aim to sort this out by separating creativity into two macro-levels: functionalist and ontological creativity.

\subsection{A Functionalist Approach} \label{functionalist}

According to Runco and Jaeger’s standard definition \cite{runco2012standard}, creativity requires \textit{originality} and \textit{effectiveness}. Ideas or products must be unusual, novel, or unique, but also useful or appropriate. Whatever name they take, these two dimensions, one requiring divergence from conventions and one requiring fitness to utility or value functions, are always present in any creativity definition and represent the two pillars around which studies in traditional computational creativity have  been developed \cite{ritchie2007some,wiggins2006searching}. A similar dichotomy is also present in Boden’s definitions of creativity, which span from the ability to generate \textit{novel} and \textit{valuable} ideas \cite{boden2009computer} to the widely known tripartite definition of creativity as the ability to come up with ideas or artefacts that are \textit{new}, \textit{surprising}, and \textit{valuable} \cite{boden2003creative}.

While surprise is still a form of divergence, it brings to the discussion the role of timeliness and causality: something creative should be unexpected and non-obvious. As a confirmation of its relevance, according to most patent laws, an invention to qualify for protection must not only be useful in some practical sense (i.e., valuable) and not previously disclosed (i.e., novel), but also not obvious to someone ordinarily skilled in that field \cite{barton2003nonobviousness}. Similarly, a painting that is a mere variation of other artworks, and that remains within the realm of standard, reasonable paintings, is less surprising and thus less creative than one that introduces disruptive elements of originality. This suggests that creativity is not a binary property but rather something that can take different forms depending on the dimension of surprise. In particular, Boden identifies three forms of creativity. \textit{Combinational} creativity involves making unfamiliar combinations of familiar ideas. \textit{Exploratory} creativity involves the exploration of conceptual spaces to come up with new ideas within them that fit with them, where a conceptual space can be seen as the generative system underlying a domain that defines a certain range of possibilities \cite{boden1994what}. \textit{Transformational} creativity involves the transformation of such spaces by generating ideas that change the pre-existing style, so that previously inconceivable thoughts become possible \cite{boden2003creative}.

The appealing nature of these creativity definitions, especially from the perspective of an AI researcher, is that they leverage observable properties. While, of course, there are plenty of examples in history of creative ideas not immediately recognised as such, effectiveness, originality, novelty, surprise, and value all concern external aspects of the creation: we just need the output (in the form of a concrete artefact or an abstract idea) to evaluate creativity under these dimensions, and similarly, we just need the output to identify what form of creativity we are dealing with. After all, observers can only evaluate the creativity of the final product and usually care only about how they feel about it. From a materialistic perspective, it is perfectly reasonable to find something creative if it appears to us as valuable (effective, appropriate, useful, or correct), novel (as something unconventional or, better, something we reasonably believe to be unconventional for the author), and to some degree surprising.

In the context of LLM-based agentic systems, we can have such a functionalist creativity at two different levels: in the final product, i.e., the outcome the agent returns to the user, and in the single actions performed, i.e., the single steps the agent makes to arrive at the final product. It is possible to have creativity at one level while not having it at the other: the final product might be creative even if the single steps are somehow standard (this can especially happen with combinational creativity, where the single elements and ideas are common and it is only their final combination that is creative), and conversely, the final product might not be creative even if some of the single steps are creative (e.g., because their final impact is negligible; a chef may invent a new way to cut carrots, but this may not impact the stew they are going to serve). Of course, it is also likely that a creative product is caused by at least one creative action and that a creative action leads to a creative product (this is particularly true for transformational creativity, where single creative actions that break norms are required).

Discussing creativity in single actions might seem complicated, as their specific effects may not be easily detectable. However, it is not really different from discussing creativity in LLMs alone: they still choose how to act by predicting the most likely action name as the completion of the provided prompt, rather than deviating from norms. In other words, they strictly follow the generative system they possess for that domain. Unless fine-tuning has pushed them (directly or indirectly) to diverge from the learned human model, they will not be able to choose an action that breaks from previous understanding of the subject, even though they can navigate its boundaries and produce something combinational or exploratory.

Evaluating creativity in the final product seems easier, as it is apparent to the user and can thus be perceived as novel, surprising, and valuable. However, its degree of creativity is influenced by the individual actions and the ``reasoning'' behind them, as these are the aspects that shape future prompts and can therefore cause it to diverge more or less from the current space of solutions. The fact that they use predetermined tools, which of course adhere to our current conception of the domain, suggests that they cannot approach transformational creativity. However, recent agents are also capable of generating new tools \cite{nguyen2025dynasaur}. In theory, this means that they can also create actions that significantly transform the conceptual space, leading to the purest forms of creativity. However, in practice, these tools are generated by an LLM, which is still a probabilistic model of human language, in a single action (whose creativity is discussed above), suggesting that it cannot generate something that deviates from what already exists in that way. This leads to a seeming conundrum: to overcome human limits, agents are provided with the ability to generate their own actions, but then they are constrained by being probabilistic models of human language and by sampling from within it, and thus remain subject to the same human limits (with the effect that they can only complement what the developer has programmed, filling a gap in the space rather than transforming it).

Indeed, LLM-based agentic systems are already capable of producing results that are valuable, novel, and surprising. However, these results are the product of interpolation (looking between the seen examples) or extrapolation (looking beyond the seen examples), but not yet of hyperpolation (transcending the seen examples) \cite{ord2024interpolation}, which is instead required to abstract from the defining dimensions of a domain and be able to alter or drop one or more of them, which is the essence of transformational creativity \cite{boden2009computer}. Similarly, these results are usually obtained via induction or deduction rather than abduction. Current LLMs cannot jump or break boundaries: they can execute the necessary steps to prove new theorems from established premises, but they are incapable of formulating original premises from sensed experience \cite{zahavy2026llms}. Again, creative abduction (the form of abduction where the law is invented \textit{ex novo} to fit with the evidence in a more elegant and even aesthetically pleasant way \cite{eco1983horns}) is conducive to transformational creativity, as it is evident in revolutionary, path-breaking scientific discoveries \cite{bertilsson2004elementary,kuhn1962structure}.

The mere fact that LLM agents cannot approach transformational creativity does not impact how creatively (in a combinational and exploratory sense) they can be, especially in scientific discovery. For instance, GPT-5.2 has recently derived a new formula in theoretical physics by spotting a pattern behind autonomously reduced expressions and positing a general formula for them \cite{guevara2026single}; a different version of the same model has also been capable of reaching the same formula and producing a formal proof by just ``reasoning'' through the problem \cite{lupsasca2026gpt}. Still, these results are closer to inductive than abductive reasoning, and to extrapolation than hyperpolation. How to effectively test for transformational creativity is an open, and potentially unsolvable, problem. Although tests such as superspace extrapolation exist, where training examples in an $n$-dimensional space are extended to an $n + 1$-dimensional superspace encompassing them \cite{lucas2012superspace}, it is hard to imagine how they could be systematically applied to the continuously expanding array of AI applications. Designing these tasks and specifying the corresponding $n + 1$-dimensional superspaces already constitutes a highly creative, transformational challenge. Most likely, the purest form of creativity can only be assessed as an emergent property, e.g., because agents start showing the ability to invent a new programming paradigm, initiate an unprecedented artistic style, derive a novel non-Euclidean geometry, or achieve any other paradigm-changing result.

Finally, until now, we have partially neglected the role of the ``reasoning'' step, focusing more on the action selection and the final output. However, this step can play a significant role from a creativity perspective. For now, the most creative and least obvious outputs from LLMs and LLM agents are due to careful prompting: there is a strong correlation between output creativity and prompt creativity, and without a sufficiently specific and originality-facilitating prompt, the output tends to collapse to slops \cite{hoel2026bits}. This is a known aspect of creativity and innovation, especially in science. As Einstein and Infeld put it, ``the formulation of a problem is often more essential than its solution, which may be merely a matter of mathematical or experimental skill. To raise new questions, new possibilities, to regard old problems from a new angle requires creative imagination and marks real advances in science'' \cite{einstein1938growth}. Whether multiple, possibly divergent ``reasoning'' steps can create the conditions for a sufficiently specific and original prompt that leads to more creative outputs may represent the first and most ideal gate toward greater functionalist creativity.

\subsection{An Ontological Approach} \label{philosophical}

While the properties of the creative output and its effect on beholders and users are all that matter from a materialistic point of view, they represent only one side of the coin in creativity theories. Indeed, there is broad agreement among researchers that creativity should be studied and evaluated from perspectives beyond the mere product (e.g., \cite{mackinnon1970creativity,sternberg1991investment,tardif1988what}).

As summarised by Rhodes \cite{rhodes1961analysis}, three additional perspectives require attention: the process, the press, and the person.
The creative \textit{process} encompasses motivation, perception, learning, thinking, and communication \cite{rhodes1961analysis}, and requires both domain- and creativity-relevant skills \cite{amabile1983social}. Typically, once a problem is provided (by external or internal stimuli), the process involves a loop where one or multiple responses are generated and evaluated; if one response is found to be successful, the loop ends; otherwise, other responses are generated based on these findings (or the problem is revisited) \cite{amabile1983social}. On the other hand, the creative \textit{press} refers to the relationship between a product and the influence its environment has upon it \cite{rhodes1961analysis}. As we already discussed, products have to be accepted as creative by society; however, they should also be influenced by previously accepted works and, in turn, influence future ones. Thus, the environment should have a key role in how the product is shaped: it provides a culturally defined domain of action in which innovation is possible, and it contains a set of peers that evaluate whether the product is worthy of being promoted and preserved \cite{csikszentmihalyi1988society}. Finally, the \textit{person} perspective acknowledges the personality, intellect, temperament, habits, attitude, value systems, and defence mechanisms of the producer \cite{rhodes1961analysis}, and requires the agent to exhibit intentionality \cite{runco2025updating} and purposes \cite{gaut2010philosophy}\footnote{An interesting dimension of creativity, as proposed by Shanahan \textit{et al.} in  \cite{shanahan2023role}, lies in the concept of role-playing. In this approach, artificial agents simulate different personas, perspectives, or narrative roles, enabling them to explore creative possibilities from multiple angles. While this mechanism can enhance AI’s capacity to generate novel ideas or solutions, it still fundamentally relies on human input, whether in defining the roles, setting the context, or guiding the objectives of the exercise. Going beyond this type of framing, in which the role-playing is driven by the AI system itself through intrinsic motivation, is an open area, which we are going to discuss in the next section.}.
We have previously demonstrated that LLMs do not satisfy these requirements \cite{franceschelli2025creativity}. The question remains: do agentic systems perform any better?


Embedding the LLM as the core part of larger systems, as those discussed in Section \ref{llm_agents}, has arguably addressed two of the known limitations of LLMs under creativity theory. First, the agentic system can now evaluate its own output and decide whether to return it or not (e.g., \cite{feng2026autonomous,knuth2026claude}). The model outputs the final solution only when it is convinced of it; otherwise, it can continue to gather new evidence from the environment, or to correct its approach via a different reasoning or the use of alternative tools. While, as already discussed, this planning and acting might not be creativity-oriented and not based on creativity-relevant skills, this \textit{generate-then-evaluate} loop closely resembles the creative process described by Amabile \cite{amabile1983social}. Second, LLM-based agentic systems can perceive and interact with the environment in ways that go beyond the mere prompt engineering typical of LLMs (even though the LLM still receives such information via automated prompts). In this way, they can acknowledge other agents, be influenced by their outputs, and deal with an evolving environment \cite{nisioti2024collective}, addressing concerns regarding the lack of a social dimension in artificial creativity.

While these two properties are partially hard-coded and limited in scope, they nevertheless address the majority of the ``easy problems'' in AI creativity highlighted in \cite{franceschelli2025creativity}, i.e., issues concerning exploration of different solutions and the production of outputs that diverge from conventions in constrained contexts. However, they still fail to resolve the ``hard problems'', which involve questions of deeper understanding, intentionality, and genuine transformation of the underlying conceptual space.
In particular, it is possible to spot three main gaps, one for each creative perspective beyond product. First, they are not intrinsically motivated and work only on external inputs. However, motivation is an essential part of the creative process and usually comes from an intrinsic interest in the task (i.e., the task is interesting or enjoyable per se) \cite{deci1985intrinsic}, which also positively impacts the creativity of discovered solutions \cite{amabile1985motivation}. Similarly, solving discovered rather than externally presented problems is more conducive to creativity \cite{getzels1975problem,runco1988problem}. Second, the influence of the environment and the impact of their own outcomes and other external products are typically limited to the specific episode and do not leave a trace on the LLM. While RAG-based systems can collect information about what has happened after the LLM deployment and solve its knowledge cut-off \cite{gao2024retrieval}, and some agentic systems allow for the generation and preservation of new, artificial tools to expand the available toolset \cite{nguyen2025dynasaur}, such an updated information is still prompted to the same, old LLM, that will process it according to the same, old probability model of the human language. In summary, LLM-based agentic systems still lack the never-ending loop in which past experience shapes future experience; they still lack continual learning.
Third, agentic systems do not possess personal traits and are not intentional agents. They lack both \textit{liberty} (independence from controlling principles) and \textit{agency} (capacity for intentional action) \cite{issak2021artistic}; indeed, the system passively follows the programmed routine, and while the agent has the autonomy of deciding the next action, it cannot refuse to generate a response. At each timestep, the LLM output is the mere product of a probabilistic model upon an automatically formed input, and the final solution is the mere output at the last timestep, rather than the result of an intentional, personal, and experience-defined process.

All in all, there is a \textit{fil rouge} linking these problems: the current lack of consciousness and self-awareness in artificial intelligence \cite{butlin2023consciousness}, despite evidence of minimal introspective capabilities \cite{binder2025looking,comsa2025does}. Indeed, being conscious is central for intentionality: to be intentional, a state or process must be thinkable or experienceable; and to be thinkable or experienceable, it must be, at least in principle, \textit{consciously} thinkable or experienceable \cite{searle1991consciousness}. In addition, it has been recently argued that continual learning is necessary (but possibly not sufficient) for consciousness, and the lack of consciousness in LLMs may be due to the lack of continual learning \cite{hoel2026disproof}. Finally, intrinsically motivated actions should be volitional and experienced as congruent and self-endorsed \cite{deci2015being}.
In other words, intelligence and sentience are two very distinct properties \cite{birch2024edge,lavelle2020machine}. While the first may be sufficient to \textit{do} something creative, both are necessary to \textit{be} creative, at least under more rigorous, stringent philosophical definitions.

\section{The Future of AI Creative Agents: To Infinity and Beyond}

While current agentic systems may fail to approach creativity under certain perspectives and may be partially limited under others, their creative achievements in recent years have been remarkable.  For example, LLMs have moved from struggling with basic maths \cite{nogueira2021investigating} to passing university-level exams \cite{borges2024could}, and now, having been embedded in larger agentic systems, they can even participate in scientific research \cite{schmidgall2025agent}. At this pace, it seems plausible to predict a future in which agents can reach the highest forms of functionalist creativity and may even be considered capable of ontological creativity. However, before proceeding down these paths, we should ask ourselves: where should we aim? Is a never-ending progression toward fully human-like creativity desirable after all?

\subsection{Transformational AI Creativity as Augmentation, Not Substitution
}
Historically, reaching transformational (functionalist) creativity does not doom current solutions to disappearance; rather, the new $n+1$-dimensional space typically allows novel and past styles of thinking to coexist. The birth of abstract painting did not cause figurative art to disappear; non-Euclidean geometries did not prevent Euclidean geometry from remaining the most studied; new programming paradigms and languages did not completely replace previous ones. On the contrary, transformationally creative artefacts can be inspirational sources for other creative outputs (as the modified space can now be explored in different ways, thus being conducive to more creativity \cite{boden2003creative}), and they might even unlock new understandings of life and science that were previously subject to traditional assumptions, as happened several times in history, from Copernican heliocentrism to Einstein's general theory of relativity. In this sense, agentic systems might, for example, overcome constraints dictated by human (physical or mental) limitations.

Moreover, functionalist creativity can also lead to outcomes that are orthogonal or complementary to our creativity. For instance, Nature is, in a sense, functionally creative: it produces artefacts (such as shells, leaves, plants, even landscapes) that can be perceived as valuable, novel, and surprising, without being \textit{truly} creative from an agentic perspective (as its underlying process is non-agentic \cite{paul2018attributing}). Artificial creativity might reach similar levels of utility in terms of aesthetics, profitability, and even research: just like the study of Nature's artefacts reveals something about the world we live in, the study of creative AI artefacts can reveal something about AI and humans as well.

In addition, moving generative AI and agentic systems from combinational (or no creativity at all) to deeper forms of creativity, where the outputs are less derivative and more divergent, can reasonably position AI outputs beside rather than in place of human outputs. Indeed, creators and researchers might use creative agents to expand and complement their work, while non-creative, replicative systems may be seen as shortcuts (in terms of time and cost) to human results. Returning less derivative products and generating more uncommon outputs may therefore reduce ethical and legal risks, preserving human intellectual property \cite{cooper2025files} as well as the established human roles in creative and innovative domains \cite{abumusab2024generative}.

Because of these considerations, we argue that pursuing functionalist creativity is a promising direction for next-generation LLM-based agentic systems. Encouraging greater, yet valuable, divergence at both the learning and inference levels, alongside enhancing reasoning and tool use with creative capabilities, can represent an important first step forward. Since hyperpolation may be reached via abduction or, conversely, may serve as a mathematical model for explaining abduction \cite{ord2024interpolation}, addressing one may also help resolve the other. Surpassing benchmarks and test suites that favour repetitive and derivative outputs, and developing new evaluative approaches, is another timely research direction.

\subsection{The Challenging Nature of Intrinsic Motivation in AI Creativity
}

Reaching ontological creativity would require a different set of skills and properties to be possessed by LLM-based agentic systems. In particular, in Section \ref{philosophical} we identify three main gaps of current agentic systems under the ontological approach: lack of intrinsic motivation; lack of experience-based continual learning; and lack of intentionality and personality. 

Intrinsic motivation is not a novel topic in AI research, especially around reinforcement learning \cite{singh2004intrinsically}; several methods already exist to incorporate a notion of intrinsic motivation and curiosity during learning, to push the agent to learn deeper and further \cite{ladosz2022exploration}. However, this is an approximate, synthetic version of intrinsic motivation, in which agents are merely incentivised to seek solutions that not only maximise environmental rewards but also possess desirable characteristics (e.g., those that favour learning or cover under-explored regions of the solution space \cite{gottlieb2013information}). The sort of (intrinsic) motivation considered by creativity theories is instead about having an intrinsic interest in the task at hand, or even choosing and reframing the task to find it more enjoyable \cite{crutchfield1962conformity}. The approximated version already studied in RL can undoubtedly have a positive impact on agentic LLMs, helping them find multiple ways of solving a problem or pushing them to explore multiple strategies during reasoning. 

Having systems that can reframe problems into something more interesting, in the sense of being challenging but still feasible, can enhance research and innovation in several ways. On the other hand, the most authentic form of intrinsic motivation may have disruptive consequences, as an enjoyable task can also be useless or too divergent from what we are asking for, potentially allowing agents to escape human control.

\subsection{Continual Learning and Self-Improvement in Creative AI Agents}

Continual learning has been receiving increasing attention from AI researchers \cite{hadsell2020embracing}. A central challenge lies in incorporating novel information while correcting or updating prior knowledge, a topic that has become a major focus of current research. Most existing approaches concentrate on extending training to new, online data while preserving previously acquired skills \cite{shi2025continual}. However, advancing beyond the knowledge contained in the original training set remains a significant and rapidly growing area of investigation\footnote{Buzz Lightyear’s iconic catchphrase ``To infinity and beyond" \cite{toystory1995} was selected as the title for this section, because it offers a compelling metaphor for the potential of AI creativity based on intrinsic motivation and continual learning. Just as the character aspires to transcend known limits of space, AI systems can push the boundaries of imagination, generating ideas, concepts, and artefacts that often surpass conventional human expectations. This represents one of the most interesting current research directions in this space.}.

As of now, the learning scheme remains fixed, and the same goes for the parameters that govern it. As a recent alternative, self-improving agents leverage the self-referential nature of these systems, which can be provided with tools to analyse, modify, and evaluate themselves, and can continually improve and simultaneously learn how to improve \cite{zhang2026hyperagents}; however, they are limited to computable tasks that allow for empirical validation (which is arguably hard in creative domains), and base their improvements exclusively upon such evaluation. On the other hand, human-like continual learning is always influenced by personal experience, and each new piece of information is acquired differently depending on its emotional impact, individual interests, and internal goals. The narrower version of continual learning can positively affect human creativity, as it can help LLM agents develop more effective interaction styles, align with user preferences, and build coherent artistic and research trajectories, in which each new creation evolves from previous ones. However, an agent that actively learns through a range of different schemes, selecting the most appropriate for each new experience, and that can set its own goals and develop its own reward functions, may diverge too far and cease to be a useful tool for creativity.

While humans rely on intrinsically motivated, experience-driven learning shaped by emotion, embodiment, and long-term goals, existing LLM agents operate through optimisation processes, reinforcement signals, and externally defined objectives. Currently, their behaviour reflects increasingly sophisticated forms of goal-directed optimisation and adaptation.
However, the current research on more autonomous, tool-using, and self-improving systems suggests that, as models gain the ability to select learning strategies, update their behaviour over time, and pursue complex objectives, they may exhibit forms of functional autonomy that reduce direct human control. Empirical studies already show that LLM agents can match or exceed human performance in specific knowledge-intensive tasks, including coding, design ideation, and scientific problem-solving, particularly when augmented with retrieval, planning, and feedback mechanisms \cite{chen2021evaluating}.

\subsection{Multi-Agent AI Systems and Creativity}

Multi-agent AI systems introduce a further dimension to computational creativity by enabling interaction, collaboration, and competition among multiple autonomous agents \cite{saunders2015computational}. Rather than treating creativity as an isolated property of a single model, we can consider it as an emergent phenomenon arising from distributed processes, where diverse agents contribute distinct perspectives, strategies, and generative biases. In creative settings, interacting agents can facilitate idea generation through constructive brainstorming and debate. Agents may propose, critique, refine, or combine outputs, leading to iterative improvement beyond what a single system could typically achieve \cite{lin2025creativity}. This is particularly relevant in tasks requiring exploration of large or structured solution spaces, where diversity of thought and possibly contrasting hypothesis generation are beneficial. In fact, in such contexts, creativity may emerge not only from individual generation but also from negotiation, coordination across and even competition among agents.

Moreover, competitive multi-agent setups can stimulate creative divergence. When agents are incentivised to outperform one another under shared or partially conflicting objectives, they may explore more unconventional or high-risk solutions \cite{zhang2026llmbased}. Cooperative settings, in contrast, tend to promote convergence and refinement, potentially improving coherence and usability of outputs \cite{li2023camel}. The balance between cooperation and competition therefore plays a key role in shaping the nature of generated artefacts \cite{duenezguzman2023social}. Indeed, future progress in AI creativity may depend not only on improving individual models, but also on designing effective ecosystems of interacting AI agents.

\subsection{AI Agentic Systems and the Transformation of Creative Work
}
Given their scalability and computational efficiency, such systems could significantly impact creative and knowledge-based professions. Economic analyses of automation and generative AI adoption indicate potential displacement or transformation of roles in art, research, and invention, alongside productivity gains \cite{acemoglu2018artificial, acemoglu2025simple, acemoglu2026ai}. At the same time, psychological research highlights the importance of creativity for identity, meaning, and well-being, suggesting that reduced human centrality in creative processes could have non-trivial psychological effects. In other words, this shift could carry enormous economic (i.e., occupational) implications, alongside significant psychological effects stemming from a diminished sense of control over what has long been considered a defining human trait: the ability to create \cite{bergson1911creative}.


\section{Conclusion}

LLMs, and especially LLM agents, are increasingly employed by researchers and practitioners to tackle a wide range of tasks traditionally associated with creativity. Yet, the question of whether these systems are genuinely creative remains contentious, hinging on how creativity is defined and the perspective from which it is assessed. In this paper, we have proposed a dualistic framework designed to reconcile these differing viewpoints and provide a principled foundation for evaluating creativity in both LLMs and LLM-based agentic systems. In particular, we have conceptualised creativity at two macro levels: functionalist creativity, which pertains to the observable properties of generated artefacts or ideas, and ontological creativity, which concerns the underlying characteristics of the generative process as well as the personal and social conditions necessary for creative acts. We have then applied these definitions to examine current agentic LLMs, arguing that they exhibit functionalist creativity to some extent but still lack the hyperpolation and abductive reasoning necessary for genuine transformational creativity. Furthermore, while embedding LLMs within larger, agentic systems endows them with certain capacities relevant to ontological creativity, they remain deficient in three fundamental areas: intrinsic motivation, continual learning, and intentionality and authenticity.

Although current capabilities remain limited, recent advances indicate that many of these constraints could be overcome in the coming years. Some of these forthcoming developments (particularly those enhancing functionalist creativity) may serve as powerful tools for human creators. Conversely, other advances could pose significant risks to creative practice. We argue that fostering discussions around creativity is essential for shaping the future of AI in creative domains across legal \cite{cooper2025files}, ethical \cite{zhu2024exploring}, and scientific \cite{lu2026rethinking} dimensions, providing a shared framework to determine both the directions we should pursue and the strategies needed to safeguard the central role of \textit{homo faber} \cite{bergson1911creative}.

\bibliography{biblio.bib}

\end{document}